\documentclass[prd,twocolumn,nofootinbib,preprintnumbers,amsmath,amssymb]{revtex4-2}

\usepackage{graphicx}
\usepackage{dcolumn}
\usepackage{bm}
\usepackage[usenames]{color}
\usepackage{color}
\usepackage{graphicx,epsfig}

\usepackage[colorlinks={true}]{hyperref}
\hypersetup{colorlinks=true,linkcolor=red,citecolor=blue,urlcolor=blue}
\usepackage{orcidlink}

\begin{document}
\title{
Comment on ``Controlling the dynamical evolution of quantum coherence and quantum correlations in $e^{+} e^{-} \rightarrow \Lambda \bar{\Lambda}$ processes at BESIII''
}
\author{Saeed~Haddadi\orcidlink{0000-0002-1596-0763}}\email{haddadi@ipm.ir}
\affiliation{School of Particles and Accelerators, Institute for Research in Fundamental Sciences (IPM), P.O. Box 19395-5531, Tehran, Iran}

\date{\today}

\begin{abstract}
We critically examine recent claims [\href{https://doi.org/10.1103/kxw9-wdth}{Phys. Rev. D {\bf 113}, 016024 (2026)}] regarding quantum coherence, steering, and non-Markovian dynamics in the hyperon–antihyperon system produced in the process $e^{+} e^{-} \rightarrow \Lambda \bar{\Lambda}$. We argue that the theoretical framework employed in the analyzed work suffers from fundamental physical inconsistencies. In particular, the treatment of the $\Lambda \bar{\Lambda}$ pair as a bipartite system evolving under correlated quantum channels is not physically justified, since the produced hyperons are free, unstable particles that do not interact with a common environment after production. Consequently, the application of open quantum system techniques, including Markovian and non-Markovian quantum channels, lacks a clear physical basis. Moreover, we show that the computation and interpretation of quantum steering for this system is operationally and conceptually meaningless, as no well-defined measurement-induced state update or controllable local measurement scenario exists for unstable relativistic particles. These issues call into question the physical relevance of the reported quantum correlations, their hierarchy, and their dynamical behavior. Our analysis highlights the necessity of carefully distinguishing between formal mathematical quantifiers of quantumness and physically realizable quantum information protocols in high-energy particle systems.
\end{abstract}

\medskip 


\maketitle

The authors~\cite{Ref1} have investigated quantum coherence and various forms of quantum correlations in the hyperon–antihyperon system produced in the process $e^{+} e^{-} \rightarrow \Lambda \bar{\Lambda}$, with particular emphasis on their angular dependence and on the role of Markovian and non-Markovian quantum channels. By employing tools from quantum information theory, they interpret the observed spin correlations as quantum resources such as coherence, entanglement, and quantum steering.

We will raise the problem that the physical assumptions underlying this analysis are not compatible with the nature of the $\Lambda \bar{\Lambda}$ system. In particular, the treatment of the produced hyperon–antihyperon pair as an open bipartite system evolving under correlated quantum channels, as well as the interpretation of quantum steering and non-Markovian dynamics in this context, lack a clear physical and operational basis.

\textbf{Remark~1.} The hyperon–antihyperon pair produced in the reaction
$
e^{+} e^{-} \rightarrow \Lambda \bar{\Lambda}
$
is generated in a single scattering event and subsequently propagates as two free, unstable, relativistic particles~\cite{Ref2}. After production, the particles do not remain embedded in a controllable environment nor do they interact with a shared bath. Therefore, modeling their post-production evolution using correlated quantum channels, as is customary in open quantum systems and quantum information theory~\cite{Ref3}, is physically unjustified. In standard quantum information settings, correlated channels arise from a common environment, engineered system–bath interactions, or explicit dynamical couplings between subsystems~\cite{Ref4}. None of these conditions apply to the $\Lambda \bar{\Lambda}$ system after its creation. As a result, the assumption that classical correlations can delay the decay of quantum coherence or quantum correlations lacks a clear physical mechanism.

\textbf{Remark~2.} The introduction of Markovian and non-Markovian regimes in the discussed work appears to be purely formal. As we know, non-Markovianity is a property of open-system dynamics, defined through information backflow between a system and its environment. However, the hyperon–antihyperon system does not interact with a structured environment. The decay of hyperons is governed by intrinsic weak-interaction lifetimes, not environmental decoherence. Indeed, there is no physically meaningful memory kernel or bath correlation function associated with the system. Consequently, classifying the system's behavior as Markovian or non-Markovian is conceptually misleading.

\textbf{Remark~3.} Quantum steering is an operationally defined form of quantum correlation. Its physical meaning relies on the ability of one party (Alice) to perform local measurements, a well-defined state-update rule conditioned on Alice's outcomes, and a controllable scenario in which Bob's conditional states can be verified. None of these requirements are satisfied for the $\Lambda \bar{\Lambda}$ system, namely:
  $i$) measurements on hyperons are indirect, inferred from decay products,
  $ii$) there is no freedom to choose measurement settings in real time,
  and $iii$) no operational steering protocol can be implemented.
Therefore, calculating steering measures for this system is not merely experimentally challenging, it is conceptually ill-defined. The resulting steering values do not correspond to any physically realizable quantum information task and should not be interpreted as evidence of steerability.

\textbf{Remark~4.} The paper~\cite{Ref1} conflates mathematical quantifiers of quantumness (such as geometric discord or coherence measures) with physical quantum resources. While it is mathematically permissible to compute these quantities from a reconstructed density matrix, their physical interpretation requires a valid operational framework.
In high-energy particle physics, density matrices often encode spin correlations fixed at production.
These correlations do not represent dynamically evolving quantum resources.
No quantum control, storage, or information processing is possible.
Thus, the reported hierarchy among quantum steering, entanglement of formation, discord, and coherence lacks operational meaning in this context.

\textbf{Main point.} The essential problem is not whether the $\Lambda\bar{\Lambda}$ pair experiences interactions during hadronization or detection, but whether these interactions can legitimately be modeled as a correlated quantum channel acting on a bipartite spin system.
Hadronization is a highly nonperturbative QCD process that creates the baryons; it is not an external dynamical map acting on a pre-existing two-qubit state. Likewise, detector interactions occur after the quantum state has already been fixed by the production process and cannot be represented as a controlled, time-continuous channel acting on a shared two-particle density matrix.
Therefore, treating the $\Lambda\bar{\Lambda}$ pair as an open bipartite system evolving under a tunable correlated noise channel lacks physical meaning. The spin density matrix extracted from decay distributions is a static statistical object conditioned on production kinematics, not a dynamical quantum state undergoing engineered decoherence. Consequently, all conclusions drawn from varying an abstract noise parameter in such a channel are physically unsupported.

In summary, the central physical assumptions underlying the analyzed work~\cite{Ref1} are not supported by the nature of the $\Lambda \bar{\Lambda}$ system. The use of correlated quantum channels, non-Markovian dynamics, and quantum steering lacks operational and physical justification. As a result, the reported calculations, interpretations, and conclusions are incorrect and should be reconsidered with greater attention to the physical constraints of elementary-particle systems.




\end{document}